\documentclass[12pt, twocolumn,showpacs,preprintnumbers,aps,pre,reprint,groupedaddress]{revtex4-1}

\usepackage{graphicx}
\usepackage{subfigure}
\usepackage{color}
\usepackage{amsmath}
\usepackage[linktocpage,colorlinks=true,linkcolor=blue,citecolor=blue,breaklinks=true]{hyperref}
\usepackage{verbatim}
\usepackage{breakcites}
\newcommand{\ju}[1]{\textcolor{red}{{\it#1}} }

\newcommand{\bl}[1]{\textcolor{blue}{#1}}
\usepackage{ulem}

\begin{document}

\preprint{}

\title{Active nematic materials with substrate friction}

\author{Sumesh P. Thampi}
\affiliation{The Rudolf Peierls Centre for Theoretical Physics, 1 Keble Road, Oxford, OX1 3NP, UK}

\author{Ramin Golestanian}
\affiliation{The Rudolf Peierls Centre for Theoretical Physics, 1 Keble Road, Oxford, OX1 3NP, UK}

\author{Julia M. Yeomans}
\affiliation{The Rudolf Peierls Centre for Theoretical Physics, 1 Keble Road, Oxford, OX1 3NP, UK}
\email[]{j.yeomans1@physics.ox.ac.uk}
\homepage[]{http://www-thphys.physics.ox.ac.uk/people/JuliaYeomans/}


\date{\today}

\begin{abstract}
{Active turbulence in dense active systems is characterised by
high vorticity on a length scale large compared to that of individual entities.
We describe the properties of active turbulence as momentum propagation is
screened by frictional damping. As friction is increased, the spacing between the walls in the nematic director field decreases as a consequence of the more rapid velocity decays. This leads to, firstly, a regime with more walls and an increased number of topological defects, and then to a jammed state in which the walls deliminate bands of opposing flow, analogous to the shear bands observed in passive complex fluids.}
\end{abstract}



\maketitle

Active materials intrinsically exist out of equilibrium driven by a continuous input of energy. Many such systems are of biological origin, including suspensions of cytoskeletal elements, cells and bacteria. Vibrated granular rods and active colloids provide examples of inanimate  active materials \cite{Sriram2010, *Ganesh2011, Marchetti2013}. Dense active systems often exhibit collective behaviour producing long range spatial and temporal correlations that are much larger than the typical length and time scales associated with individual particles \cite{julia2012, Aranson2012, dogic2012, Poujade2007, Petitjean2010,Narayan2007}. Active turbulence, observed in active nematics, is a canonical example of this behaviour, which is characterised by rapidly varying flow fields with a high degree of vorticity.
Recently it has been proposed that there are links between active turbulence and the appearance of lines of kinks, or walls, in the director field and the formation of topological defects also appears to be relevant in defining the active turbulent state \cite{dogic2012, Giomi2013, ourprl2013, ourepl2014, Igor2013, Shelly2014}.

 Active turbulence has been observed in experiments on systems as diverse as dense suspensions of bacteria \cite{julia2012, *Aranson2012}, microtubules and molecular motors \cite{dogic2012}, cells  \cite{Poujade2007, *Petitjean2010},  mixtures of bacteria and liquid crystals \cite{Igor2013}, and vertically vibrated rods  \cite{Narayan2007}.  Diverse simulations from those considering discrete particles \cite{Graham2005, *Shelly2012, *Shi2013, Shelly2014}, eg driven, self-avoiding rods of high aspect ratio, to those based on differing continuum theories, eg active nematohydrodynamics, \cite{Jorn2013, ourprl2013, Giomi2013} all see the hallmarks of active turbulence. However we still lack a predictive theory of this dynamical state, or an understanding of which of the properties of active turbulence are universal and which are system specific.


A classification often applied to active nematics is the relevance of momentum conservation in determining their dynamics \cite{Marchetti2013}. Systems that conserve momentum (eg a free-standing fluid layer) are termed `wet', those with no momentum conservation (eg vibrated granular rods)  `dry'. In reality completely wet or dry systems are two ends of a continuum.
 We are not aware of any analysis investigating connections between these limits, a question important in establishing the extent to which active turbulence is universal. Therefore, the aim of this paper is to study active turbulence in the crossover regime between wet and dry nematics when momentum propagation is increasingly being damped.

We stress that understanding the crossover between wet and dry systems is not just a theoretical curiosity but is also of relevance to almost all experiments performed so far in active nematics. This is because these experiments are performed in two dimensions with the system in contact with a solid or fluid  substrate which will lead to some degree of frictional damping. Experiments using microtubules and molecular motors are conducted on an oil-water interface \cite{dogic2012} and so frictional damping may be relatively small,
pattern formation studies of amoeboid cells \cite{Kremkemer2000}, collective migration \cite{Poujade2007} and structured flows \cite{Lenepreprint} of epithelial cells are performed on a solid substrate and therefore both viscous and frictional damping will be relevant, whereas for vertically vibrated granular rods \cite{Narayan2007} friction will dominate. Comparison of theory with experiments is crucial in guiding the development of the theory of active turbulence. Moreover, to match coefficients in the continuum theories to physical systems, the role of friction must be understood.


We shall show that substrate friction promotes shear banding. Consequently, the number density of walls and topological defects in the director field increases, this density controlled by the strength of friction coefficient compared to viscosity. 
This significantly changes the structure and characteristic length scales of active turbulence. High vorticity flow fields, which evolve rapidly in time, observed at small to moderate friction, are replaced by a dynamical steady state of bands of alternating shear at large friction.
We first illustrate this concept in one dimension, and then present the results of simulations demonstrating its consequences in two dimensions.

We consider, along with equation of continuity, the hydrodynamic equations of motion for an active nematic modified by an additional frictional term \cite{sriram1982, sriram2003}, 
\begin{align}
\rho (\partial_t + u_k \partial_k) u_i = \partial_j \Pi_{ij} - \gamma u_i ,
\label{eqn:ns} \\
(\partial_t + u_k \partial_k) Q_{ij} - S_{ij} = \Gamma H_{ij}
\label{eqn:lc}
\end{align}
where $\gamma$ is the friction coefficient. The evolution of orientational order is described by a tensor order parameter $\mathbf{Q}$, familiar in standard liquid crystal hydrodynamics \cite{DeGennesBook, *Berisbook}.  The generalised advection term,
$S_{ij} = (\lambda E_{ik} + \Omega_{ik})(Q_{kj} + \delta_{kj}/3) + (Q_{ik} + \delta_{ik}/3)
 (\lambda E_{kj} - \Omega_{kj}) - 2 \lambda (Q_{ij} + \delta_{ij}/3)(Q_{kl}\partial_k u_l)$,
describes the response of $\mathbf{Q}$ to velocity gradients. Here $E_{ij} = (\partial_i u_j + \partial_j u_i)/2$ is the strain rate tensor, $\Omega_{ij} = (\partial_j u_i - \partial_i u_j)/2$ is the vorticity tensor and $\lambda$ is the alignment parameter. We use $\lambda=0.7$, which corresponds to aligning rods \cite{DeGennesBook, *Berisbook, Scott2009}. The rotational diffusivity is denoted as $\Gamma$ and the Landau-de Gennes free energy, 
$\mathcal{F} = K (\partial_k Q_{ij})^2/2 + A Q_{ij} Q_{ji}/2 + B Q_{ij} Q_{jk} Q_{ki}/3 + C (Q_{ij} Q_{ji})^2/4,$
%
governs the relaxation of the orientational order by determining the molecular field $H_{ij} = -\delta \mathcal{F}/ \delta Q_{ij} + (\delta_{ij}/3) {\rm Tr} (\delta \mathcal{F}/ \delta Q_{kl})$ in Eq.~(\ref{eqn:lc}). $A, B$ and $C$ are material constants and $K$ is the elastic constant. 

The standard, passive liquid crystal contributions to the stress are the viscous stress, $\Pi_{ij}^{viscous} = 2 \mu E_{ij}$,  
and the elastic stress,
%
%
$\Pi_{ij}^{passive}=-P\delta_{ij} + 2 \lambda(Q_{ij} + \delta_{ij}/3) (Q_{kl} H_{lk})
-\lambda H_{ik} (Q_{kj} + \delta_{kj}/3)  - \lambda (Q_{ik} + \delta_{ik}/3) H_{kj}
-\partial_i Q_{kl} \frac{\delta \mathcal{F}}{\delta \partial_j Q_{lk}} + Q_{ik}H_{kj} - H_{ik} Q_{kj}$
where $P$ is the pressure. The active contribution to the stress is a consequence of the dipolar nature of the active system, $\Pi_{ij}^{active} = -\zeta Q_{ij}$ \cite{Sriram2002}. Thus any gradient in $\mathbf{Q}$ generates a flow field and $\zeta$, the activity coefficient, determines the strength of this flow field. We consider $\zeta>0$, corresponding to extensile systems. Details of the model can be found in  \cite{DeGennesBook, *Berisbook, Denniston2004, Davide2007, *Suzanne2011,ourpta2014}. We do not have density fluctuations and neglect curvature induced active currents of density commonly used in describing dry active nematics \cite{Marchetti2013}.


The numerical solution of Eqs.~(\ref{eqn:ns}) and (\ref{eqn:lc}) uses a hybrid lattice Boltzmann method on a $D3Q15$ lattice \cite{Davide2007, *Suzanne2011} in the incompressible limit. The parameters used are $\Gamma=0.34$, $A=0.0$, $B=-0.3$, $C=0.3$, $K=0.01$, $\zeta=0.01$, $\mu=2/3$, in lattice units, unless mentioned otherwise. Results for different values of A, corresponding to different temperatures, are shown in the appendix \ref{appsec:aneq0}.  As usual discrete space and time steps are chosen as unity and therefore all quantities can be converted to physical units appropriately depending on the material of interest
 \cite{Cates2008, *Henrich2010, ourprl2013}. The frictional force $\gamma \mathbf{u}$ was incorporated into the lattice Boltzmann scheme as a force density in a way analogous to eg. gravity \cite{Guo2002, Davide2007, *Suzanne2011}.

\begin{figure}
\subfigure[]{\includegraphics[trim = 0 0 0 0, clip, width=\linewidth]{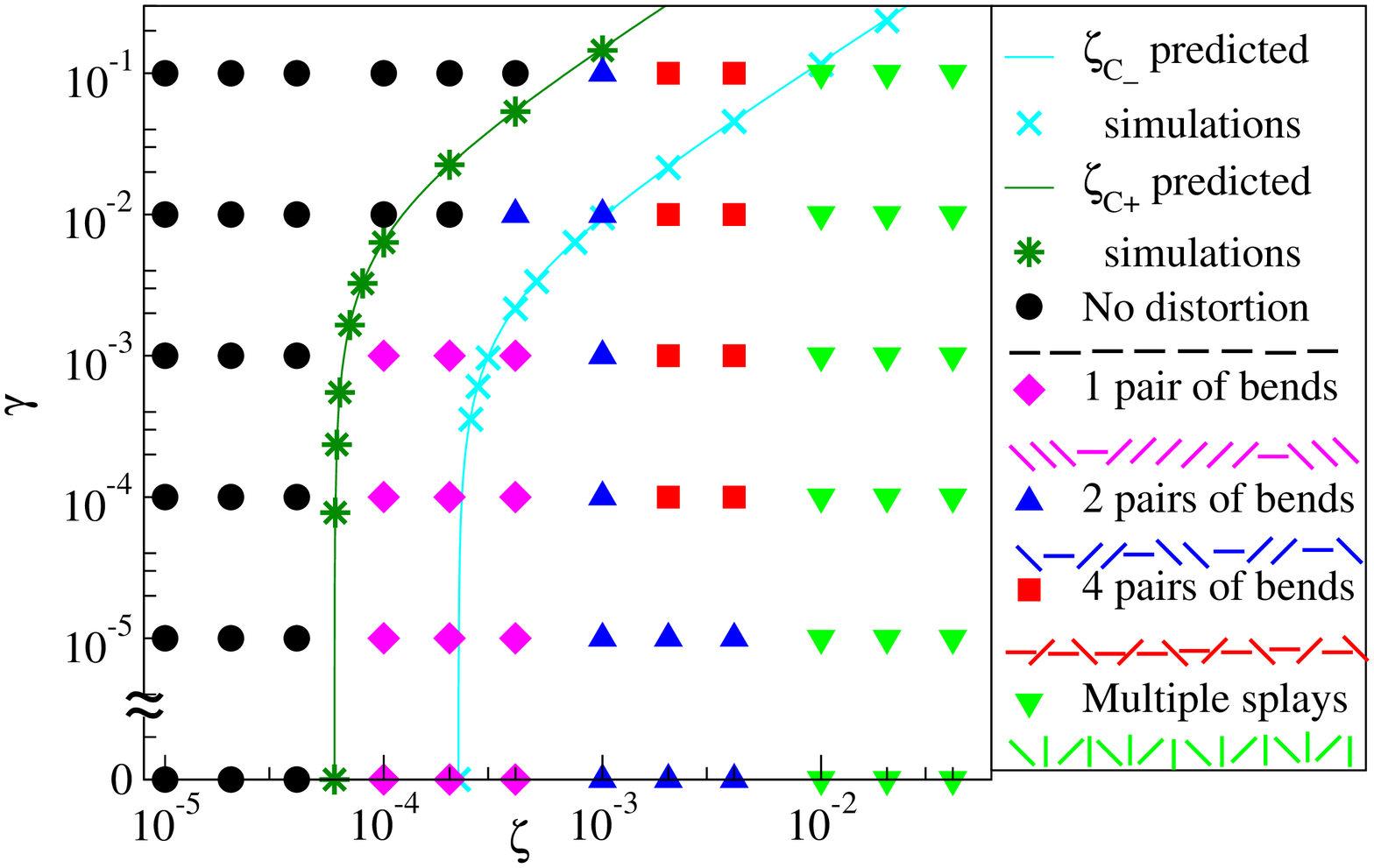}\label{fig:pbc}}
\subfigure[]{\includegraphics[trim = 0 0 30 80, clip, width=\linewidth]{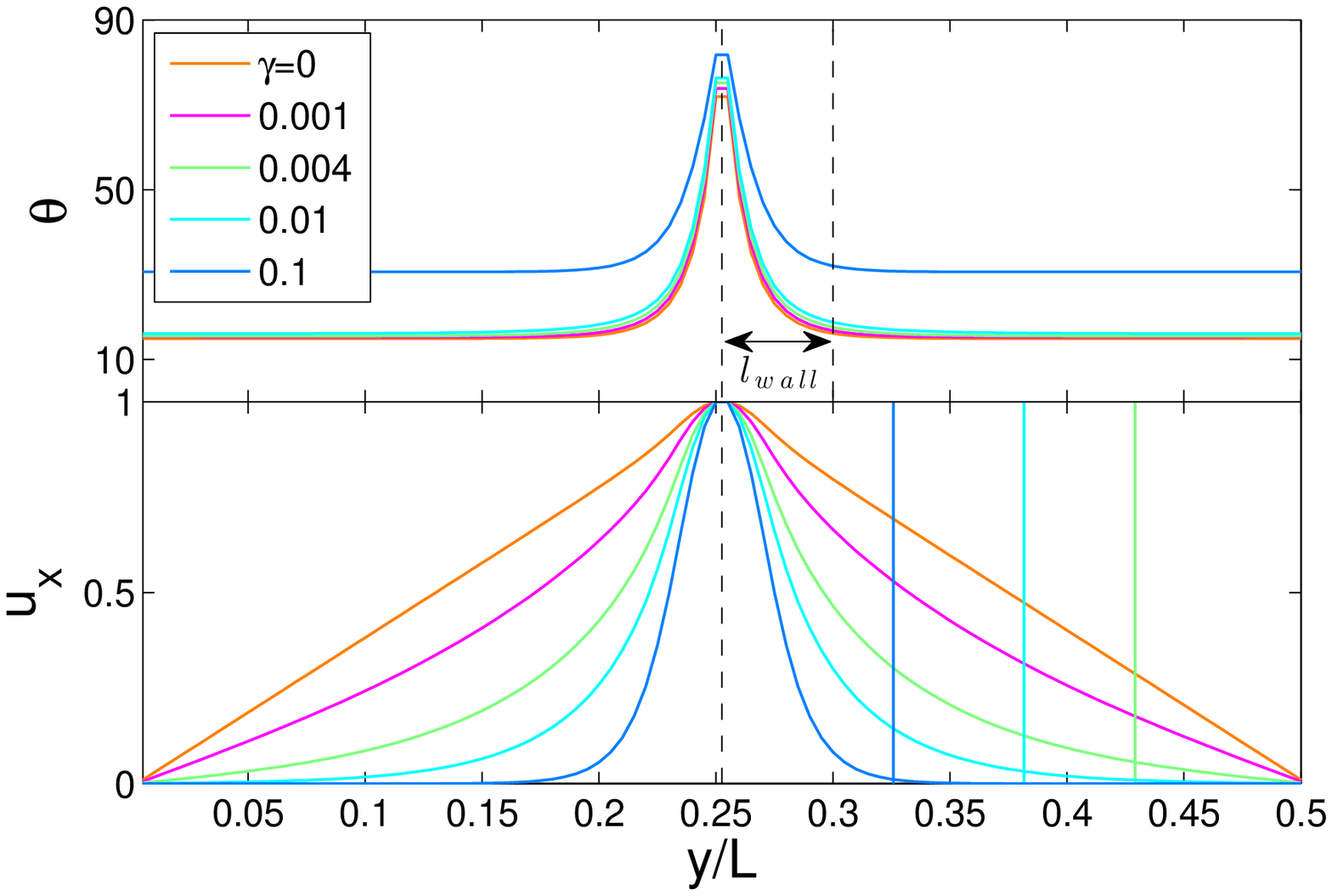}\label{fig:veldir}}
\caption{(Color online) (a) Steady state director field configurations for different values of activity $\zeta$ and friction coefficient $\gamma$ for a 1D system with periodic boundary conditions. The increase in thresholds of $\zeta_{C_\mp}$ is calculated analytically as a function of $\gamma$ and is plotted as continuous lines. The symbols on these lines represent the thresholds obtained numerically by simulating the full equations Eq.~\ref{eqn:ns}-\ref{eqn:lc}. (b) Director angle measured with respect to the flow direction (top) and magnitude of the velocity field associated with a single bend deformation  (bottom). The vertical continuous lines, drawn at $l_{wall}+2l_{diss}$, show the screening effects on the velocity field.}
\label{fig:phase}
\end{figure}

Before presenting results for the full 2D model it is instructive to consider 1D solutions, confining the active nematic between $y=0$ and $y=L=100$, with periodic boundary conditions along $y$, and assuming translational symmetry in the $x$ direction.  The nematic field can now be represented by a single variable, $\theta(y)$ describing the angle between the direction of orientation and the $x$ axis. A consequence of the active stress is that gradients in $\theta(y)$ generate a velocity  in the $x$ direction $u_x(y)$. This system shows strong hysteresis, with the final state depending on the initial conditions \cite{Davide2007B}. Therefore we chose the protocol of always initialising the simulation with four pairs of oppositely oriented bend deformations along $y$ (see legend in Fig.~\ref{fig:pbc}). It relaxed to a variety of different steady states depending on the values of the friction $\gamma$ and the activity $\zeta$, as summarised in  Fig.~\ref{fig:pbc}.  Other initial conditions give qualitatively similar results.

At low activities, the system regains a nematic configuration, with the director field aligned along $y$, and no flow.  Using a mathematical analogy to the  Fredericks transition in passive nematic liquid crystals Voituriez {\it et al.} \cite{joanny2005} showed that, for $\gamma=0$, the nematic ordering is unstable to a state with spontaneous flow at a critical activity $\zeta^{wet}_{C_\mp}=\pi^2K[4\Gamma_1\mu+(\lambda_1\mp1)^2]/[2L^2(\lambda_1\mp1)]$. Here $\mp$ denotes an instability driven by small deviations from initial angles $\theta=0$ and $\theta=\frac{\pi}{2}$ respectively. Also $\Gamma_1 = 2 \Gamma/9q^2$ and $\lambda_1 = (3q+4)\lambda/9q$ where $q$ is the magnitude of the nematic order, the largest eigenvalue of $\mathbf{Q}$ \cite{Davide2007}. The approach in \cite{joanny2005} generalises to non-zero $\gamma$ by replacing $\zeta$ by an effective activity  $\zeta-\frac{2\gamma\Gamma_1 K}{\lambda_1\mp1}$. For flow aligning nematics ($\lambda_1>1$), the threshold of activity required for the spontaneous flow transition is therefore increased due to the presence of frictional dissipation to $\zeta_{C_\mp} = \zeta^{wet}_{C_\mp} + \frac{2\gamma\Gamma_1K}{\lambda_1\mp1}$. These increased thresholds of activity as a function of friction coefficient are plotted as continuous lines in activity-friction plane in Fig.~\ref{fig:phase}. The symbols on these lines represent the thresholds obtained numerically by simulating the full equations Eq.~\ref{eqn:ns}-\ref{eqn:lc}. There is no discernible difference between theory and simulation.

For small $\gamma$ the analytical and  numerical results  show that the spontaneous flow transition is to a state where the director field  exhibits a pair of oppositely oriented bend deformations, symmetrically placed in the domain, as illustrated in Fig.~\ref{fig:pbc}. This bidirectional flow, referred to as shear banding, is a consequence of unstable regions in the stress-strain curve for active nematics \cite{Cates2008, Suzanne2011}. As activity is increased, the number of such bend deformations increases, until at $\zeta \sim 0.01$ they are no longer stable, and are replaced by splay deformations (see Fig.~\ref{fig:pbc}) as observed in \cite{ourepl2014}. As $\gamma$ increases the critical activity required to drive the transition to a flowing state also increases. Moreover the number of pairs of oppositely oriented bend deformations increases for a given activity.

To motivate why higher frictional dissipation leads to a reduction in width of the shear bands   Fig.~\ref{fig:veldir} shows the director angle, $\theta$,  (top) and the associated velocity field, $u_x$, (bottom) corresponding to a single bend deformation. Variation in the director field, localised around $y/L=0.25$, generates active forces and acts to drive the flow \cite{Madan2007, Scott2009}. When $\gamma=0$  the source flow decays slowly with a constant shear rate due to viscous damping, giving a characteristic triangular flow profile. As $\gamma$ increases the additional friction accelerates the decay of the velocity field. 

Considering a balance between the dissipative terms and modelling the source of flow generation as a point force of strength $F$ located at $y=y_0$, we see that $\mu \frac{d^2u_x}{dy^2} - \gamma u_x = F \delta(y-y_0)$. This indicates an exponential relaxation of velocity over a characteristic  length $l_{diss}=\sqrt{\mu/\gamma}$. {For small friction this length scale is not relevant,} but at large $\gamma$, it can be small so that momentum is lost over much shorter lengths than expected from viscosity alone. This is explicitly demonstrated in Fig.~\ref{fig:veldir} where  the continuous vertical lines show  $l_{wall}+~2 l_{diss}$ indicating the faster decay of velocity field with increasing $\gamma$. Here $l_{wall} $ is a measure of the width of the bend deformation itself which, as shown in the figure, is very weakly dependent on the friction.


\begin{figure}
\includegraphics[trim = 0 0 0 0, clip, width = 0.9\linewidth]{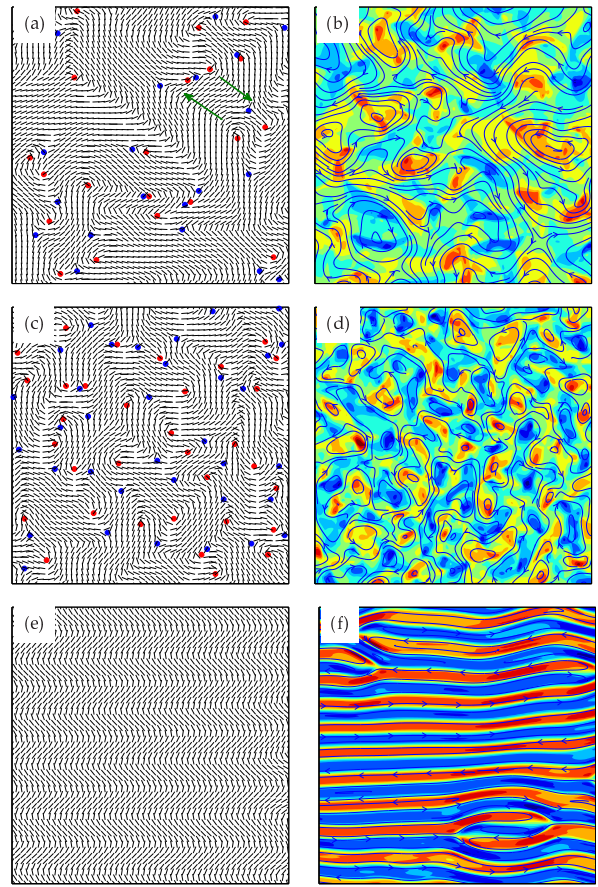}
\caption{(Color online) Left panel: director field and topological defects (with $+1/2$ and $-1/2$ defects denoted by red and blue dots) and right panel: vorticity, denoted by the colour scale from +ve (red) to -ve (blue) normalised to its maximum value, and velocity field (arrows). (a),(b): no solid friction ($\gamma=0$). (c),(d): $\gamma=0.01$ and (e),(f):  $\gamma=0.5$. (a),(c),(e) correspond to the left bottom quarter of (b),(d),(f) respectively. The formation of a pair of walls is identified in (a).}
\label{fig:twod}
\end{figure}

\begin{figure*}
\includegraphics[trim = 0 0 0 0, clip, width = \linewidth]{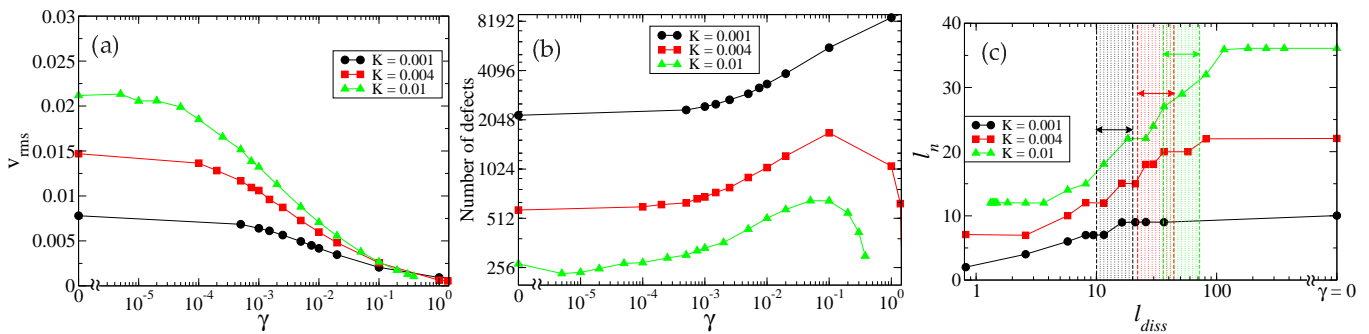}
\caption{(Color online) (a) The  RMS velocity decreases and (b) the defect density increases as the substrate friction, $\gamma$ increases. At very large $\gamma$ parallel walls become more stable with decreasing velocity and fewer defect creation events. (c) The distance between the walls, $l_n$ decreases when it becomes comparable to the dissipation screening length  $l_{diss}=\sqrt{\mu/\gamma}$. The shaded area between $l_n^{wet}$ and $2 l_n^{wet}$ shows the transition region where the screening effects start to dominate.}
\label{fig:numbers}
\end{figure*}

This shortfall of viscous momentum transfer in the transverse direction supports the formation of more closely spaced shear bands. This represents a transition towards the dry active limit where long ranged hydrodynamic interactions are absent. Similar effects are important in fully developed active turbulence. To demonstrate this
we now present the results of simulations on a two dimensional domain of size $400 \times 400$ with periodic boundary conditions, a system that exhibits active turbulence. For no friction any nematic region is unstable to the active stress. The instability leads to the formations of walls, lines of kinks in the director field.  Generally the walls are generated in pairs (as indicated in Fig.~\ref{fig:twod}(a)). They are analogous to the bend deformations observed in 1D but, unlike in 1D, they are unstable to both transverse fluctuations and, for sufficiently large activities, to the  generation of pairs of topological defects, which move apart and hence locally restore nematic order. Typical configuartions of director field and defects, and of velocity and vorticity fields, are illustrated in Fig.~\ref{fig:twod}(a) and (b) respectively.

Panels (c) and (d) of Fig.~\ref{fig:twod} compare similar snapshots, but for $\gamma=0.01$. Just as for shear-banding in the 1D systems, the addition of friction reduces the spacing between walls resulting in a larger number of walls, and thus defects, in the steady state. Distortions in the director field act as sources of flow and therefore walls and defects are the main sources of vorticity. Since the number of such distortions is increased, the flow field also changes substantially. This is apparent when comparing Fig.~\ref{fig:twod}(b) and (d): the decrease in the characteristic length of circulating regions is noticeable. 

The very different  behaviour of the active nematic for large friction, $\gamma=0.5$, is shown in the snapshots in panels (e) and (f) of Fig.~\ref{fig:twod}. The induced flow velocities are smaller and no defects are generated. Instead one observes closely spaced walls jammed approximately parallel to each other (but with some quenched disorder) supporting anti-parallel shearing flows very reminiscent of the multiple bend configurations observed in 1D. Reducing $l_{diss}$ further decreased the width of the shear bands to an unphysical limit set by lattice spacing. 
We comment that walls which do not decay into defects are also observed in active turbulence at $\gamma=0$ for sufficiently small activities. However these are highly unstable to lateral perturbations; they continually collide and show chaotic and transient behaviour \cite{ourprl2013}. 
%

We next present quantitative data to illustrate the crossover between viscous and frictional damping. 
Fig.~\ref{fig:numbers}(a) shows the root mean square velocity of the active turbulent flow as a function of the friction coefficient. Since the characteristic velocity and length scales of active turbulence for $\gamma=0$ are highly dependent on elasticity \cite{ourpta2014}, we present results for three different values of the elastic constant. As expected, the velocity decreases with increasing friction, and at large values of friction coefficient it is substantially reduced. 

Fig.~\ref{fig:numbers}(b) plots the number of defects as a function of the friction coefficient for each value of $K$. This first increases with friction, as the number of walls which finally decay into defects increases, and then decreases as there is not enough energy available to create defects for high frictional damping. The maximum occurs for smaller $\gamma$ at larger $K$ because the defect energy increases with increasing elastic constant.

In 1D we argued that the distance between walls (or equivalently the width of the shear bands) is regulated by the dissipation length $l_{diss}=\sqrt{\mu/\gamma}$ which controls the decay of the velocity field. We now present evidence that the same is true in the 2D case. We measured the order parameter correlation function
 $C_{QQ}(|\mathbf{R}'-\mathbf{R}|) = (\langle\mathbf{Q} (\mathbf{R}):\mathbf{Q}(\mathbf{R}')\rangle$ and 
estimated the distance between walls $l_n$ as the point at which $C_{QQ}$ reaches its minimum value. This is justified because  $C_{QQ}$ measures the size of nematically oriented regions. In an earlier paper we showed that $l_n$ is set by the density of topological defects \cite{ourpta2014} when $\gamma=0$. We now plot this distance, \bl{$l_n$} as a function of the dissipation length in Fig.~\ref{fig:numbers}(c). As can be seen, $l_n$ is essentially independent of the friction $\gamma$ as long as $l_n < l_{diss}$. Once these two quantities become comparable, the distance between the walls is also controlled by $l_{diss}$ and therefore decreases with $\gamma$. The region between $l_n^{wet}$ (i.e $l_n$ for $\gamma=0$) and $2l_n^{wet}$ which is highlighted in each case corresponds to the transition to the over damped limit.


To summarise, we describe here the implications of frictional damping due, for example, to flow over a substrate, in a wet active nematic. The dissipation length, $l_{diss}$, the characteristic length scale describing the decay of momentum from a source, is irrelevant for small friction. However for larger friction this length scale controls the spacing between walls. $ l_{diss}$ decreases with increasing friction and therefore the number of walls and the number of defects formed by their decay, increases. At higher friction, as the walls become even more closely spaced, they form a jammed state where wall fluctuations and break up, and hence defect formation, are suppressed. In this limit the walls are reminiscent of the boundaries between velocity bands seen in sheared passive fluids. It will be interesting to test the consequences of momentum screening in experiments by using  substrates with varying friction coefficients, and to consider whether similar effects are seen in compressible active nematics such as vibrated granular rods where density fluctuations are relevant.


\begin{acknowledgments}
We acknowledge funding from the ERC Advanced Grant
MiCE. We also thank the KITP at UCSB for hospitality and
acknowledge funding by NSF grant PHY11-25915.
\end{acknowledgments}

\appendix

\section{Choice of parameters in the free energy.}
\label{appsec:aneq0}

In \cite{ourpta2014}, we studied active turbulence in the absence of friction and performed a detailed analysis of the effect of various parameters such as activity ($\zeta$),  elasticity ($K$) and rotational diffusion constant ($\Gamma$).  Activity and elasticity had an effect on the characteristic length scales but not on the qualitative nature of the active turbulence and we expect that varying these parameters will similarly not give qualitative changes in behaviour when friction is present. Additionally, we now show that the behaviour in the presence of friction is independent of the choice of parameters in free energy. The temperature of the equilibrium nematic phase is varied by changing the coefficient $A$. The results below are for simulations at two different temperatures to those in Fig.~\ref{fig:twod}.

\begin{figure*}
\includegraphics[trim = 0 0 0 0, clip, width = 0.9\linewidth]{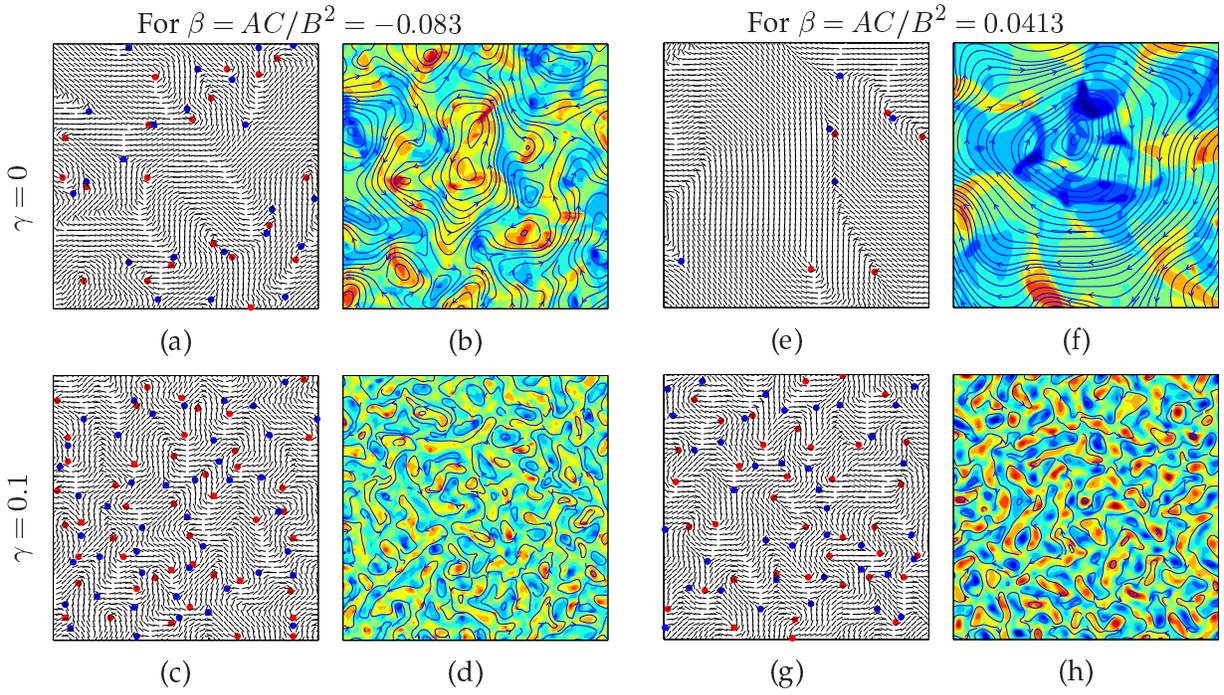}
\caption{(Color online) Effect of friction on the director field (and topological defects) and flow field (both vorticity and streamlines) at two different temperatures are illustrated here. On left hand side (panels (a)-(d)), $A=-0.025$ or, equivalently, $\beta=AC/B^2 = -0.083$ corresponding to a temperature, $T < T^*$ where $T^*$ is the temperature close to isotropic nematic transition temperature $T_{NI}$, about which the free energy is expanded. Below $T^*$ the isotropic phase is completely unstable. On the right hand side (panels (e)-(h)), $A=0.0124$ or $\beta=AC/B^2 = 0.0413$ corresponding to a temperature, $T$ very close to $T^{**}$ where $T^{**}$ is the highest temperature at which the nematic phase exists. Above $T^{**}$ the nematic phase is completely unstable.}
\label{fig:temp}
\end{figure*}

In Fig.~\ref{fig:temp}, both the director field with topological defects and the corresponding vorticity field with streamlines are shown. The first row (panels a,b and e,f)  are simulations with no substrate ($\gamma=0$). We obtain active turbulence at these two widely differing temperatures with a quantitative difference that the number of defects decreases with increasing temperature. The second row (panels c,d and g,h) shows the results of introducing a substrate by increasing the friction coefficient to $\gamma=0.1$. We get an increased number of defects and smaller vorticity structures as the dissipation length comes into play.

It is worth mentioning that values or even the range of free energy parameters or the temperature in these active systems are not known.  The fact that friction introduces a new length scale will be important in trying to extract parameters like $A$ by comparing with experiments.

\bibliography{refe.bib}

\end{document}